\documentclass[pra,superscriptaddress,twocolumn]{revtex4}

\usepackage[dvips]{graphicx}%
\usepackage{bm,color}
\usepackage{ulem}
\usepackage{amsmath,amssymb}
\newcommand{\braket}[2]{\langle #1 | #2 \rangle}
\newcommand{\ket}[1]{\left |  #1 \right \rangle}
\newcommand{\bra}[1]{ \left \langle #1  \right |}
\newcommand{\ave}[1]{  \langle #1  \rangle}
\newcommand{\ketbra}[2]{\ket{#1}\bra{#2}} 
\def \tr{{\textrm {Tr}}}

\begin{document}

\title{Schmidt-number benchmark for genuine quantum memories and gates}

\author{Ryo Namiki}\affiliation{Department of Physics, Graduate School of Science, Kyoto University, Kyoto 606-8502, Japan}

\author{Yuuki Tokunaga}\affiliation{NTT Information Sharing Platform Laboratories, NTT Corporation, 3-9-11 Midori-cho, Musashino, Tokyo, 180-8585, Japan}\affiliation{Japan Science and Technology Agency, CREST, 5 Sanban-cho, Chiyoda-ku, Tokyo, 102-0075, Japan}

\date{\today}

\begin{abstract}
%

We propose to apply the notion of the Schmidt number in order to show that a quantum memory or gate process is capable of maintaining a genuine multi-level quantum coherence. We present a simple criterion in terms of an average gate fidelity with respect to the input states of two mutually unbiased bases   
 and demonstrate  the coherence of the gate operation 
 in several existing experiments. The Schmidt-number benchmark naturally includes the quantum benchmark as a specific case. 
 \end{abstract}

\maketitle

 
Quantum memories and gates are important elements for future quantum information 
processing such as quantum computing and quantum communication network 
\cite{Lvovsky2009,Hammerer2010,Ladd2010}. 
They are required to maintain quantum coherence as well as classical logical information.
An important theoretical task is to establish a clear benchmark for 
the maintenance of quantum coherence or proof of entanglement associated with the experimental approaches \cite{Pop94,Bra00,Horo99}. 

The quantum benchmark gives a basic experimental success criterion for quantum memories and gates \cite{Ham05,Rig06,Takano08,namiki07,namiki08,Has08,Has10,Fuc03}. In order to estimate the performance of the memory
or gate process, an average of the input-output fidelities
is often measured. The quantum benchmark fidelity is defined as the limit of the
average fidelity achieved by entanglement breaking (EB) maps \cite{HSR2002}. If the measured fidelity surpasses the benchmark fidelity, the process 
 can maintain the coherence of an  entangled signal.

Recently, not only single-qubit processes
\cite{Chen2008,Neeley2008,Olmschenk2009} but also multi-qubit or 
multi-dimensional quantum processes have been experimentally 
investigated such as the storage of entangled states
\cite{Choi2008, Akiba2009} or two-qubit controlled-NOT gates
\cite{O'Brien2004, Okamoto2005, Bao2007}.
In general, a legitimate quantum process for a $d$-level quantum (qudit) system is
expected to keep a set of coherent superpositions of $d$-orthogonal vectors. 
 Otherwise, it degrades entanglement of correlated signals 
 and quantum information processing
cannot be performed faithfully. 
Therefore, it is important to verify the maintenance of higher dimensional
quantum coherence for such a multi-dimensional process. 

The Schmidt number is as a simple measure of entanglement that represents  how many degrees of freedom are involved in entangled states \cite{Ter20}. It was utilized to demonstrate the generation of a genuine multi-dimensional entanglement \cite{Tokunaga08,Inoue09}. 
Hence, it is natural to invoke the notion of the Schmidt number in order to demonstrate a genuine multi-level coherence of the physical process as well. 

In this Letter, we generalize the quantum benchmark and present a simple criterion for verifying a genuine multi-level quantum coherence of the quantum channel acting on a $d$-level system
by using 2$d$ input states of two mutually unbiased bases. 
From this criterion the coherence of the gate operation in several existing experiments is verified.
The experimental test requires the number of input-output measurement setup $2d^2$, which is substantially smaller than the number of $d^4$ needed for   the process tomography.

Any physical process is described by a completely positive
trace-preserving map (quantum channel) \cite{NC00}, 
namely, the process $\mathcal E$ satisfies the properties (i) $\tr \mathcal E (\rho ) =1 $ for any density operator $\rho \ge 0$ and $\tr \rho =1$,  and (ii) the positivity of the total density operator is maintained if the process acts on the subsystem $A$ of any combined system $AB$, i.e., $ \mathcal E_A \otimes I _B (\rho _{AB}) \ge 0$.
A natural isomorphism \cite{Jam} of the quantum channel 
$\mathcal E$ acting on a qudit system is written by
\begin{eqnarray}
J_{\mathcal E} := \mathcal E_A \otimes I_B (\hat \Phi_{0,0}) \nonumber 
\end{eqnarray}
where $I$ represents the identity operation and $\hat \Phi_{0,0}:= \ket{\Phi_{0,0}}\bra{\Phi_{0,0}}$ is the maximally entangled state in the two-qudit system,
$|\Phi_{0,0} \rangle := \frac{1}{\sqrt d }\sum_{j=0}^{d-1} |j \rangle_A | j  \rangle _B. $  
Since, $J_{\mathcal E }$ is a bipartite quantum state, 
the concepts developed for characterizing entanglement can be also applied for the characterization of the quantum channel.

The Schmidt number for the quantum channel 
${\mathcal E}$ can be defined as the Schmidt number for $J_{\mathcal E}$ \cite{Hua06}.
We call $J_{\mathcal E}$ has Schmidt number of $k$ if it can be 
decomposed into the mixture of pure states whose Schmidt number is 
at most $k$ and there exists a pure state 
with the Schmidt number at least $k$ in any decomposition. 
The Schmidt number $k$ for the quantum channel implies that an entanglement with the Schmidt number of $k$ is left after the channel action on the subsystem of the maximally entangled system. Its physical meaning is that quantum coherence of a $k$-dimensional space can be maintained in the process.  

For the case of $k=1$, $J_{\mathcal E}$ has a separable form 
and the corresponding channel ${\mathcal E}$ is referred to as the classical measurement-and-preparation (MP) scheme 
or EB channel 
 \cite{HSR2002}. 
The separable form 
$J_{\mathcal E} =\sum_{i}\ketbra{\phi_i}{\phi_i}\otimes
\ketbra{\varphi_i}{\varphi_i}$ 
with the condition $\sum_{i}\ketbra{\varphi_i}{\varphi_i}= \openone /d$ 
directly implies the expression of the MP scheme: 
$\mathcal E (\rho ) = \sum_i d \bra{\varphi_i^*}\rho\ket{\varphi_i^*}
\ketbra{\phi_i}{\phi_i}$ where
$\ket{\varphi^*}:=\sum_{j=0}^{d-1}\braket{\varphi }{j}\ket{j} $ and the
operator $  M_i:= d \ketbra{\varphi_i^*}{\varphi_i^*}  $ forms a
positive operator valued measure $   \sum_i M_i =\openone$. 
This implies that the process returns the state based on the transfer of the classical information represented by the index $i$ of the measurement outcomes and never transmits the quantum correlation. 
On the other hand, for any process being incompatible with the classical
MP scheme, 
one can find an entangled state 
that maintains inseparability in the process. 

Hence, it is fundamental to 
eliminate the possibility that the process can be described by a MP scheme. The quantum benchmark is devoted to this EB or non-EB paradigm and concerns whether or not the channel outperforms the processes with the Schmidt number of $k=1$. However, it cannot tell us whether the channel is capable of maintaining the multi-level coherence for $k\ge 3$. 
In what follows, we make a stronger criterion for verifying that the channel 
 outperforms any channel whose Schmidt number is lower than a given Schmidt number of $k  \le d $ in terms of the average fidelity. 

We define the average fidelity of the process $\mathcal E$ 
on the transformation
task from a set of input states $\{ | \psi_i \rangle \}$ to a set of
\textit{target} states $\{| \psi_i' \rangle \}$ with respect to a prior
distribution $\{p_i\}$ by 
\begin{eqnarray}
\bar F &=& \sum _i p_i \langle \psi_i'| {\mathcal E} \left( |\psi_i \rangle\langle \psi_i |  \right) | \psi_i'\rangle. \nonumber 
\end{eqnarray} 
If the concerning process is an implementation of a unitary gate $U$, a natural choice of the target states is $\ket{\psi_i' } = U \ket{\psi_i} $ for any $i$ so that  the ideal process $\mathcal E (\rho) = U \rho U ^\dagger  $ yields the unit fidelity $\bar F = \sum _i p_i \langle \psi_i | U ^\dagger  {\mathcal E} \left( |\psi_i \rangle\langle \psi_i |  \right) U | \psi_i \rangle = \sum _i p_i |\langle \psi_i   |\psi_i \rangle | ^2=1 $.  
We define the limit of the average fidelity achieved by the process whose Schmidt number is $k$:   
\begin{eqnarray}
\bar F^{(k)} &:=& \max_{{\mathcal E} \in \mathcal O_k } \sum _i p_i \langle \psi_i'| {\mathcal E} \left( |\psi_i \rangle\langle \psi_i |  \right) | \psi_i'\rangle, \nonumber 
\end{eqnarray}
where $\mathcal O_k$ denotes the set of the channels with the Schmidt
number of $k$ or less than $k$. Note that $\bar F^{(k)} \le \bar F^{(k')} $ holds for any $k \le k'$ since $\mathcal O_k \subset \mathcal O_{k'}$.
The quantum benchmark fidelity  \cite{Pop94,Bra00,Ham05,Takano08,namiki07,namiki08,Fuc03,Has10} is defined as the lowest Schmidt-number case of $k=1$ and
corresponds to the maximization of the average fidelity over MP
schemes, 
\begin{eqnarray}
\bar F ^{(1)} &=& \max_{{\mathcal E} \in \mathcal O_1 } \sum _i p_i \langle \psi_i'| {\mathcal E} \left( |\psi_i \rangle\langle \psi_i |  \right) | \psi_i'\rangle \nonumber  \\ 
  &=&\max_{  M_k,  \phi_k} \sum_i \sum_k  p_i \langle \psi_i |  M_k |\psi_i \rangle  | \langle \psi_i' | \phi_k \rangle| ^2 .   \nonumber 
\end{eqnarray}
If the measured fidelity exceeds this classical limit, it is evident 
that the process outperforms any classical MP scheme and is capable 
of transmitting entanglement. 
Further, if the measured fidelity exceeds $\bar F^{(k)}$, we can 
eliminate the possibility that the process is simulated by the 
class of quantum channels with Schmidt number $k$ and less than $k$, namely, it can transmit an entangled state whose Schmidt number is at least $k+1$. This establishes the Schmidt-number 
 benchmark in order that 
 the process outperforms a wide class of ``weaker'' 
processes including the lowest class of the EB channels. 

  
 Note that the Schmidt number is an entanglement monotone, which is a function of quantum state monotone under the local operation and classical communication \cite{mono}. 
  Besides the Schmidt number, there are many other entanglement monotones. A wide range of entanglement monotones specifies separable states as the class of states who has the lowest degree of entanglement.  
 Hence, it is possible to employ any of such monotones on $J_{\mathcal E}$ in order to construct general benchmarks beyond the quantum benchmark. 
The degree of entanglement is invariant under local unitary operations. This property implies that the channels $\mathcal E $ and $\mathcal E ' $ are equivalently coherent if $J_{\mathcal E '} = u_A\otimes v_B J_{\mathcal E} (u_A\otimes v_B )^\dagger $ where $u$ and $v$ denote unitary operators on single qudit system. The transformation by the local unitary operators corresponds to the unitary operations before and after the process because $\mathcal E' (\rho )= u \mathcal E  ( v^\dagger \rho v )u^\dagger $ holds. 
In this classification, any unitary channel is equivalent to the identity channel.

Next, we consider the task of an identity map associated  with quantum memories and derive the fidelity limits in the case that the set of input states forms two mutually unbiased bases. 
Due to the unitary equivalence, the fidelity limits are the same if the target states are connected with the input states by a unitary operation as $\ket{\psi_i' } = U \ket{\psi_i} $ \cite{namiki07}.  

\textit{Quantum benchmark fidelity---.} 
%
An ideal qudit memory is expected to store classical 
$d$ logical bits information. Corresponding to this type of classical 
information, the memory is supposed to store at least a set of 
$d$ orthogonal states $\{\ket{j }\}$ with a high fidelity. 
For later convenience we call $\{\ket{j }\}$ $Z$-basis. 
We define the average fidelity with respect to the input of the $Z$-basis states as
\begin{eqnarray}
F_Z({\mathcal{E}}) &:=& \frac{1}{d}\sum_{j=0}^{d-1}  \bra{j}\mathcal{E}(\ket{j}\bra{j})\ket{j}. \label{defFz}
\end{eqnarray}
This fidelity represents the memory performance of the classical information storage. 
We can see that the MP scheme $\mathcal E_Z^{EB} (\rho ) = \sum_{j=0}^{d-1} \ketbra{j}{j} \rho \ketbra{j}{j}$ enables faithful storage of the classical information, i.e., 
$
F_Z({\mathcal{E}_Z^{EB}}) = 1. 
$

In addition to the storage of the classical information, the ideal quantum
memory is expected to store the coherent superpositions of the 
logical basis states, namely, the quantum memory is required to
store the coherence between the elements of the $Z$-basis states 
with a high fidelity. For this purpose, it is natural to consider 
 uniform superpositions of the $Z$-basis states and estimate 
the fidelity for such states.  
We define the $X$-basis states as a Fourier transform of the $Z$-basis states by 
$| \overline {k} \rangle :=  \frac{1}{\sqrt d} \sum_{j=0}^{d-1 }e^{i\frac{2  \pi  }{d } k j}\ket{j }$, 
and define the average fidelity with respect to the input of the $X$-basis states by
\begin{eqnarray}
F_X({\mathcal{E}}) &:=& \frac{1}{d}\sum_{j=0}^{d-1}  \bra{\overline j}\mathcal{E}(\ket{\overline j}\bra{\overline j})\ket{\overline j}. \label{defFx}
\end{eqnarray}
The classical MP scheme can achieve either $F_Z=1$ or  $F_X=1$ by
choosing the measurement of the $Z$-basis or the $X$-basis, but cannot
achieve both of the unit fidelities simultaneously.
In order to see the limit of the classical scheme as a quantum memory, it is natural to ask how high one can enlarge the average of $F_Z$ and $F_X$ without  quantum coherence. 

Let us define the average fidelity on the two conjugate bases as 
\begin{eqnarray}
F_{\mathcal{E}} &:=& \frac{1}{2}(F_Z(\mathcal{E})+F_X(\mathcal{E}) ).  
\label{maindef} \end{eqnarray} 
We can show that the limit of this average fidelity achieved by classical MP schemes is given by
\begin{eqnarray}
  F^{(1)} := \max_{\mathcal{E} \in \mathcal O_1 }F_{\mathcal{E}} &=& \frac{1}{2}(1 + \frac{1}{d}).  \label{theo1}
\end{eqnarray}

\textit{Proof. }
From 
 Eqs. (\ref{defFz}), (\ref{defFx}), and  (\ref{maindef}), we can write
\begin{eqnarray}F_{\mathcal{E}} & =& \frac{1}{2d}\sum_{j=0}^{d-1} \left( \bra{j}\mathcal{E}(\ket{j}\bra{j})\ket{j}+\bra{\overline{j}}\mathcal{E}(\ket{\overline{j}}\bra{\overline{j}})\ket{\overline{j}} \right) \nonumber \\
& =& \frac{1}{2 }  \tr\Big[  \mathcal{E}\otimes I  (\hat \Phi_{0,0} )   \nonumber \\&&\times \sum_{j=0}^{d-1}\Big(    \ket{j}\bra{j}\otimes\ket{j}\bra{j}    +  \ket{\overline{j}}\bra{\overline{j}}\otimes\ket{\overline{-j}}\bra{\overline{-j}}  \Big) \Big] \nonumber \\ & =:& \frac{1}{2 } \langle\hat C_d \rangle_{  \mathcal{E}\otimes I  (\hat \Phi_{0,0} )}, \nonumber \end{eqnarray}
where we use the expression of the maximally entangled state in the 
$Z$-basis and $X$-basis, 
$|\Phi_{0,0} \rangle = \frac{1}{\sqrt d }\sum_{j=0}^{d-1} | j \rangle |j \rangle = 
\frac{1}{\sqrt d }\sum_{j=0}^{d-1}    |\overline {  j}  \rangle  | \overline{ -j}  \rangle   $,
and define the operator of the total qudit correlation 
\begin{eqnarray}
\hat C_d :=\sum_{j=0}^{d-1}\Big(    \ket{j}\bra{j}\otimes\ket{j}\bra{j}  
  +  \ket{\overline{j}}\bra{\overline{j}}\otimes\ket{\overline{-j}}\bra{\overline{-j}}  \Big). \nonumber 
\end{eqnarray}
For any  EB channel $\mathcal E \in \mathcal O_1$, 
${  \mathcal{E}\otimes I  (\hat \Phi_{0,0} )} $ is a separable state. 
 Therefore, the maximal fidelity obtained by EB channels is no larger than the maximal expectation value of the correlation obtained by separable states,  namely, 
\begin{eqnarray}
\max_{\mathcal{E} \in \mathcal O_1 }F_{\mathcal{E}} \le \max_{\rho \in S_1} \ave{ C_d }_\rho,   \label{o1case}
\end{eqnarray}
where $S_1$ is the set of separable states. 
It is direct to see that the correlation has the Bell-diagonal expression: 
\begin{eqnarray}
 \hat C_d &=&\sum_{l=0}^{d-1}  \hat \Phi_{l,0}+\sum_{m=0}^{d-1}  \hat \Phi_{0,m} 
\label{BDC}\end{eqnarray}
where $|\Phi_{l,m} \rangle  := \hat X_A^l \hat Z_B^m  |\Phi_{0,0} \rangle $ is the Bell states with the following definition of the generalized Pauli operators,  \begin{eqnarray}
\hat Z:&=&\sum_{j= 0}^{d-1} e^{ i \frac{2 \pi}{d}  j} |j  \rangle\langle  j|, \  
\hat X: = \sum_{j=0}^{d-1}|j+1 \rangle\langle j|.  \nonumber
\end{eqnarray} 
Using Eq. (\ref{BDC}) and the completeness of the Bell basis $\sum_{l,m=0}^{d-1}  \hat \Phi_{l,m}= \openone$, we have
\begin{eqnarray}
 \hat C_d  
 &=& \openone + \hat \Phi_{0 ,0} -\sum_{l,m=1}^{d-1}  \hat \Phi_{l,m} \le  \openone + \hat \Phi_{0 ,0}. \label{sdeq}\end{eqnarray}
From this relation, Eq. (\ref{o1case}), and the well-known relation for the maximally entangled fraction $ \max_{\rho\in S_1}  \ave{\hat \Phi_{0.0}}_{\rho} = 1/d$ in Eq. (36) of Ref. \cite{Horo99-a}, 
we can obtain 
\begin{eqnarray}
 F ^{(1)} = \max_{\mathcal{E} \in  \mathcal O_1  }F_{\mathcal{E}} \le  \max_{\rho \in S_1} \ave{ C_d }_\rho  \le\frac{1}{2}(1 + \frac{1}{d}).  \nonumber
\end{eqnarray}
This bound can be achieved by the MP scheme $\mathcal E_Z^{EB} (\rho ) = \sum_{j=0}^{d-1} \ketbra{j}{j} \rho \ketbra{j}{j}$. 
This concludes Eq. (\ref{theo1}). \hfill$\blacksquare$

Consequently, we can show that the channel outperforms any EB
channel when the average fidelity is larger than $F^{(1)}$. 
This criterion can be tested by using only the number of $2d $ input elements of the two conjugate bases.  
Note that, from Eq. (\ref{sdeq}), we can see that the state is uniquely determined to be $\ket{\Phi_{0 ,0}}$ in the case of  $F_{\mathcal E} =1 $. This case of $J_{\mathcal E} = \hat \Phi_{0,0}$ implies that the process is  exactly an identical map and a perfect quantum memory. The Bell-diagonal forms of Eqs. (\ref{BDC}) and (\ref{sdeq}) suggest that $F_{\mathcal E}$ is highly affected by the noises including both $\hat X$ and $\hat  Z$, while it is rather insensitive to the noises including only $\hat X$ or $\hat  Z$.  Inequality of Eq. (\ref{sdeq}) also suggests that a lower bound of the process fidelity is given by  $F_{\mathcal E}$.   

\textit{Fidelity limit for Schmidt number of $k$.---} 
Now we consider how high one can enlarge the average fidelity $ F_{\mathcal{E}}$ with the channels whose Schmidt number is at most $k \le d$, or equivalently $\mathcal E \in \mathcal O_k$. We can show that  
\begin{eqnarray}
F^{(k)}:=  \max_{\mathcal{E} \in \mathcal O_{k } } F_{\mathcal{E}} &=& \frac{1}{2}(1 + \frac{k }{d}).  \label{2ndr}
\end{eqnarray}

\noindent\textit{Proof.---} 
Similar to the derivation of Eq. (\ref{o1case})  we have
\begin{eqnarray}
\max_{\mathcal E  \in \mathcal O _k } F_{\mathcal E }\le  \frac{1}{2 }  \max_{\rho\in S_{k}} \ave{\hat C_d}_\rho,  \nonumber 
\end{eqnarray}
where $S_k$ denotes the set of the states whose Schmidt number is at most $k$.
Then, Eq. (\ref{sdeq}) and the relation, $ \max_{\rho\in S_{k}}  \langle \hat \Phi_{0,0}\rangle_\rho\le k/d$, from  Lemma 1 of Ref. \cite{Ter20} lead to 
\begin{eqnarray}
\max_{\mathcal E  \in \mathcal O _k } F_{\mathcal E }\le  \frac{1}{2 }
 \max_{\rho\in S_{k}} \ave{\hat C_d}_\rho\le \frac{1}{2} \left(1+
							  \frac{k}{d}\right). \nonumber 
\end{eqnarray}
This inequality is saturated by the channel $\mathcal E_k( \rho):= \frac{1}{k} \sum_{l=0}^{d-1} K_l \rho K_l^\dagger 
$
with $  K_l :=\hat X^l  \sum_{m=0}^{k-1} \ket{m}\bra{m}( \hat X^\dagger )^l$.
We can confirm that $\mathcal E_k( \rho) \in \mathcal O_k$ since the Kraus operator $K_l/\sqrt{k}$ gives the Schmidt number of $k$ vectors for the corresponding state, e.g.,  $\sqrt{d}  K_0 \ket{\Phi_{0,0}}= \sum_{j=0}^{k-1} \ket{j}  \ket{j}$. 
We can also verify $\tr_A [(\mathcal E_k )_A \otimes I_B (\hat \Phi_ {0,0} ) ] = \openone_B /d $, $F_Z (\mathcal E_k ) =1 $, $F_X (\mathcal E_k ) =k/d $,  and  $F_{\mathcal E_k }= F^{(k)} $, by direct calculation. This concludes Eq. (\ref{2ndr}).  \hfill$\blacksquare$

Therefore, 
the Schmidt number of the channel is at least $k+1$ when the average
fidelity of Eq. (\ref{maindef}) is larger than the $k$-th limit 
$F^{(k)} $.
Hence, the benchmark for the full-level coherence on qudit gates is to surpass the $(d-1)$-th limit of  $F^{(d-1)} $.

\textit{Application for multi-qubit experiments.} 
 In multi-qubit experiments, non-local access to the qubits is usually limited. This suggests that experimental test of the criterion can be simpler when the input and target states are products of the local qubit-states. In this regard, we can choose the pair of two conjugate bases as products of the local-qubit $Z$ and $X$ bases, i.e.,  $\{\ket{j_N} \ket{j_{N-1}}  \cdots \ket{j_1}  \} $ and $\{\ket{\overline{ {j_N}}} \ket{\overline{ j_{N-1}}}  \cdots \ket{\overline{j_1}}  \} $ with $d= 2^N$ and $j_k\in \{0,1\}$. By choosing the input states in this manner we can assign a product set of the target states not only for the case of the identical gates but also for the important case of two-qubit controlled-NOT gates. 
Further 
 we can derive the same fidelity limits with modification of the above proof by using the products of the ordinary Pauli operators $\hat X^{l} := \sigma_x ^{l_N} \otimes \sigma_x ^{l_{N-1}} \otimes \cdots  \otimes \sigma_x ^{l_1} $ and $\hat Z^{m} := \sigma_z ^{m_N} \otimes \sigma_z ^{m_{N-1}} \otimes \cdots  \otimes \sigma_z ^{m_1} $ instead of the generalized Pauli operators where $l_k, m_k \in \{0,1\} $.  

For one-qubit memory or teleportation process, 
the fidelity limit  $F ^{(1)}$ is  0.75.
From the experiment of \cite{Olmschenk2009}, $F_{\mathcal{E}}=0.90$ is obtained.
This demonstrates that the one-qubit process outperforms any classical MP scheme.
Here, our criterion needs only half number of input-output measurement setup compared to
one-qubit process tomography.
For two-qubit controlled-NOT gate, the fidelity limits are $F ^{(1)}=0.625$, $F ^{(2)}=0.75$, and $F ^{(3)}=0.875$.    
From the experiment of \cite{Okamoto2005}, $F_{\mathcal{E}}=0.86$ is obtained.
Thus, the present criterion ensures that the demonstrated gate outperforms any channel of
Schmidt number 2, but does not ensure outperforming the channels of Schmidt number 3 (This does not deny the possibility that $J_{\mathcal E}$ has Schmidt number 4).
In \cite{Bao2007}, $F_{\mathcal{E}}=0.89$ is obtained.
In this case, the criterion ensures the full-dimensional coherence of the demonstrated two-qubit gate. 
Here the number of input-output measurement setup for two-qubit
processes is 32. 
This number is significantly smaller than 256, which is needed for the complete process tomography.

Note that the fidelity limits  for the uniform average fidelity 
 can be proven to be $\bar F ^{(k)} = \frac{1+k}{1+d }$ (the limits for the process fidelity is $ \frac{k}{d}$) \cite{NunP1}. Although the uniform average fidelity or process fidelity \cite{Horo99,Nielsen02} is a standard figure for the process estimation, it requires the number of measurement settings  the same order to $d^4$ of the process tomography,  and is rather hard to access in experiments.

In conclusion, we have proposed to apply the notion of the Schmidt number in order to verify a genuine multi-level coherence of quantum channels beyond the quantum benchmark. We have presented a simple criterion for such performance on qudit channels with the use of $2d$ input states ($2d^2$ input-output measurement setup). 
 From this criterion, we have verified the multi-level coherence in several existing experiments. 
This work will provide legitimate and convenient milestones for realizing  quantum
memories and gates.



R.N. acknowledges support from JSPS.
This work was supported by GCOE Program ``The Next Generation of Physics, Spun from Universality and Emergence'' from MEXT of Japan. 

\end{document}